\documentclass[aps,preprint]{revtex4}%
\usepackage{amsfonts}
\usepackage{amsmath}
\usepackage{amssymb}
\usepackage{graphicx}%
\providecommand{\U}[1]{\protect\rule{.1in}{.1in}}

\begin{document}
\preprint{ }
\title[Dark solitons]{Finite temperature effective field theory for dark solitons in superfluid
Fermi gases}
\author{S.N. Klimin}
\altaffiliation{Department of Theoretical Physics, State University of Moldova}

\affiliation{TQC, Universiteit Antwerpen, Universiteitsplein 1, B-2610 Antwerpen, Belgium}
\author{J. Tempere}
\altaffiliation{Lyman Laboratory of Physics, Harvard University}

\affiliation{TQC, Universiteit Antwerpen, Universiteitsplein 1, B-2610 Antwerpen, Belgium}
\author{J.T. Devreese}
\altaffiliation{Technische Universiteit Eindhoven}

\affiliation{TQC, Universiteit Antwerpen, Universiteitsplein 1, B-2610 Antwerpen, Belgium}
\pacs{PACS number}

\begin{abstract}
We use a finite temperature effective field theory recently developed for
superfluid Fermi gases to investigate the properties of dark solitons in these
superfluids. Our approach provides an analytic solution for the dip in the
order parameter and the phase profile accross the soliton, which can be
compared with results obtained in the framework of the Bogoliubov -- de Gennes
equations. We present results in the whole range of the BCS-BEC crossover, for
arbitrary temperatures, and taking into account Gaussian fluctuations about
the saddle point. The obtained analytic solutions yield an exact
energy-momentum relation for a dark soliton showing that the soliton in a
Fermi gas behaves like a classical particle even at nonzero temperatures. The
spatial profile of the pair field and for the parameters of state for the
soliton are analytically studied. In the strong-coupling regime and/or for
sufficiently high temperatures, the obtained analytic solutions match well the
numeric results obtained using the Bogoliubov -- de Gennes equations.

\end{abstract}
\date{\today}
\maketitle

\section{Introduction \label{sec:intro}}

The recent progress in the experimental and theoretical study of quantum gases
has been particularly stimulated by the fact that they represent an example of
macroscopic quantum phenomena where the system parameters can be finely tuned.
For the Fermi gases, a crossover between the Bardeen--Cooper--Schrieffer (BCS)
superfluid constituted by Cooper pairs and a Bose--Einstein condensate (BEC)
of bound fermion pairs can be achieved using the Feshbach resonance. Nonlinear
phenomena in the ultracold atomic gases, especially relating to vortices and
solitons, draw a great interest, as they provide insight into the interplay of
interactions and coherence \cite{Anderson,Davis}. A dark soliton in a
superfluid is a nonlinear solitary excitation of the order parameter which
propagates on a uniform or plane-wave background and manifests itself through
a density dip. Dark solitons are one of the first fundamental nonlinear
excitations which have been experimentally detected in BECs
\cite{Burger1999,Denschlag2000}. Besides ultracold gases, there are
experimental observations of dark solitons in other systems, including optical
\cite{Chen} and mechanical \cite{Denardo,Chabchoub} dark solitons. However,
the experimental realization of dark solitons in the atomic $^{6}$Li Fermi
gases near a Feshbach resonance occurred only recently \cite{Yefsah}. The
study of these dark solitons in Fermi gases, especially with respect to the
snake instability and the subsequent decay into vortex filaments or rings, is
a subjects of active current
debate\cite{CetoliPRA88,BulgacPRL112,WenPRA88,ToikkaPRA87}.

Dark solitons in the Bose-Einstein condensates were successfully treated
theoretically using the Gross-Pitaevskii (GP) equation \cite{SolReview}, both
within the mean-field approximation \cite{Denschlag2000,Feder2000} and taking
into account quantum and thermal fluctuations
\cite{Jackson2006,Jackson2007,Martin2010,Dziarmaga2003,Gangardt2010}. For the
theoretical description of solitons in the superfluid Fermi gases in the
BCS-BEC crossover regime, however, the GP equation appears not applicable,
except in the deep BEC limit, where the pairs can be approximately considered
as a Bose gas of molecules. For the solitons within the BCS--BEC crossover,
one of the most reliable methods is based on the Bogoliubov--de Gennes (BdG)
equations (see the review \cite{Spuntarelli2010}). However, solving the BdG
equations is computationally very demanding, due to the necessity of using a
large amount of fermionic wave functions. As a result, the analysis of the BdG
solutions for dark excitons, at least at present, has been performed only in
the zero-temperature case \cite{WenPRA88,Liao2011,Spuntarelli2011}.
Computationally less demanding extensions of the BdG approach based on
coarse-graining have been developed recently\cite{SimonucciPRB89}, but not yet
applied to solitons, because in the present form it is not time-dependent.
This inspires attempts to develop complementary approaches for the ultracold
Fermi gases exploiting only a macroscopic wave function. Prominent examples of
such attempts are the modifications of the Ginzburg -- Landau (GL) approach
\cite{deMelo1993,HuangPRA79} for cold Fermi gases in the BCS-BEC crossover,
and the Gross -- Pitaevskii (GP) and nonlinear Schr\"{o}dinger (NS) equations
\cite{Adhikari2008}. The GL method is valid in a rather narrow temperature
region close to the critical temperature $T_{c}$. The GP equation works well
in the BEC regime, but can fail at weaker couplings. There are extensions of
the GL approach to lower temperatures based on expanding the free energy in
powers of the small parameter $\eta\equiv1-T/T_{c}$
\cite{ShanenkoPRL106,Shanenko2,Sh3,Orlova} or using a microscopic treatment
\cite{BabaevPRB72,BabaevPRL105,BabaevPRB86}. Our recent investigation
\cite{ExtGL,ExtGL2}, focused on the atomic Fermi gases in the BCS-BEC
crossover regime, has been devoted to the development of an effective method
for the description of the macroscopic wave function of a fermionic superfluid
system without assuming $\eta$ small. In Refs. \cite{ExtGL,ExtGL2}, the GL
formalism has been extended to the whole temperature range below $T_{c}$ for a
multiband superfluid fermion system. In the limit $T\rightarrow T_{c}$, the
theory of Ref. \cite{deMelo1993} is retrieved.

In the present work, we apply the effective field theory (EFT) of Refs.
\cite{ExtGL} to the dark solitons in a superfluid Fermi gas with $s$-wave
pairing. The fermion system is treated in the BCS-BEC crossover and in the
whole range of temperatures below $T_{c}$. The study is performed both within
the mean-field approximation and taking into account Gaussian fluctuations in
renormalization of the chemical potential of the Fermi gas. The mean-field
results are compared with BdG data in the low-temperature limit, allowing us
to reliably establish the range of validity of the effective field theory.

The paper is organized as follows. In Sec. \ref{sec:eqs}, we present the field
equations and derive their analytic solution for dark solitons. In Sec.
\ref{sec:rels}, the macroscopic integrals of motion for the soliton are
calculated, and the exact energy-momentum relation is derived for the soliton.
Sec. \ref{sec:results} contains a discussion of the numeric results for the
parameters of the dark soliton, followed by Sec. \ref{sec:conclusions}, the Conclusions.

\section{Field equations \label{sec:eqs}}

For the analytic treatment of dark solitons in ultracold Fermi gases we use
the effective field formalism developed in Refs. \cite{ExtGL,ExtGL2}. The
description of a Fermi gas with the $s$-wave pairing within this formalism is
performed using the effective action for the pair field $\Psi(r,\tau)$ (the
macroscopic order parameter). We start from the Euclidean-time form of the
effective field action from Ref. \cite{ExtGL2},%
\begin{equation}
S\left(  \beta\right)  =\int_{0}^{\beta}d\tau\int d\mathbf{r}\left[  \frac
{D}{2}\left(  \bar{\Psi}\frac{\partial\Psi}{\partial\tau}-\frac{\partial
\bar{\Psi}}{\partial\tau}\Psi\right)  +\mathcal{H}\right]  , \label{Seucl}%
\end{equation}
where $\beta$ is the inverse to the temperature, and $\mathcal{H}$ is the
Hamiltonian of the pair field,%
\begin{equation}
\mathcal{H}=\Omega_{s}+\frac{C}{2m}\left\vert \nabla_{\mathbf{r}}%
\Psi\right\vert ^{2}-\frac{E}{2m}\left(  \nabla_{\mathbf{r}}\left\vert
\Psi\right\vert ^{2}\right)  ^{2} \label{H}%
\end{equation}
with the field-dependent thermodynamic potential,%
\begin{align}
\Omega_{s}  &  =-\int\frac{d\mathbf{k}}{\left(  2\pi\right)  ^{3}}\left[
\frac{1}{\beta}\ln\left(  2\cosh\beta E_{\mathbf{k}}+2\cosh\beta\zeta\right)
\right. \nonumber\\
&  \left.  -\xi_{\mathbf{k}}-\frac{m\left\vert \Psi\right\vert ^{2}}{k^{2}%
}\right]  -\frac{m\left\vert \Psi\right\vert ^{2}}{4\pi a_{s}}. \label{Ws}%
\end{align}
Here, $a_{s}$ is the scattering length for $s$-wave pairing, $\xi_{\mathbf{k}%
}=\frac{k^{2}}{2m}-\mu$ is the kinetic energy of the fermionic atoms with mass
$m$, measured from the chemical potential $\mu$. The present formalism can
take into account population imbalance by introducing separate chemical
potentials for \textquotedblleft spin-up\textquotedblright\ and
\textquotedblleft spin-down\textquotedblright\ atoms, combined into $\mu
\equiv\left(  \mu_{\uparrow}+\mu_{\downarrow}\right)  /2$ and $\zeta
\equiv\left(  \mu_{\uparrow}-\mu_{\downarrow}\right)  /2$ $\mu_{\uparrow}$.
Finally, $E_{\mathbf{k}}=\sqrt{\xi_{\mathbf{k}}^{2}+\left\vert \Psi\right\vert
^{2}}$ is the Bogoliubov excitation energy. Expression (\ref{Ws}) formally
coincides with the saddle-point grand-canonical thermodynamic potential for
imbalanced Fermi gases \cite{PRA2008}. The coefficients in front of the
gradients are given by:%
\begin{align}
C  &  =\int\frac{d\mathbf{k}}{\left(  2\pi\right)  ^{3}}\frac{k^{2}}{3m}%
f_{2}\left(  \beta,E_{\mathbf{k}},\zeta\right)  ,\label{c}\\
D  &  =\int\frac{d\mathbf{k}}{\left(  2\pi\right)  ^{3}}\frac{\xi_{\mathbf{k}%
}}{\left\vert \Psi\right\vert ^{2}}\left[  f_{1}\left(  \beta,\xi_{\mathbf{k}%
},\zeta\right)  -f_{1}\left(  \beta,E_{\mathbf{k}},\zeta\right)  \right]
,\label{d}\\
E  &  =2\int\frac{d\mathbf{k}}{\left(  2\pi\right)  ^{3}}\frac{k^{2}}{3m}%
\xi_{\mathbf{k}}^{2}~f_{4}\left(  \beta,E_{\mathbf{k}},\zeta\right)  .
\label{EF}%
\end{align}
The functions $f_{s}\left(  \beta,\varepsilon,\zeta\right)  $ are determined
through sums over the fermionic Matsubara frequencies $\omega_{n}=\left(
2n+1\right)  \pi/\beta$:%
\begin{equation}
f_{s}\left(  \beta,\varepsilon,\zeta\right)  \equiv\frac{1}{\beta}%
\sum_{n=-\infty}^{\infty}\frac{1}{\left[  \left(  \omega_{n}-i\zeta\right)
^{2}+\varepsilon^{2}\right]  ^{s}}. \label{msums}%
\end{equation}
For any integer $s$, these sums are analytically calculated, using the
recurrence relations given in Ref. \cite{ExtGL}:%
\begin{align}
f_{1}\left(  \beta,\varepsilon,\zeta\right)   &  =\frac{1}{2\varepsilon}%
\frac{\sinh\left(  \beta\varepsilon\right)  }{\cosh\left(  \beta
\varepsilon\right)  +\cosh\left(  \beta\zeta\right)  },\label{msum}\\
f_{s+1}\left(  \beta,\varepsilon,\zeta\right)   &  =-\frac{1}{2s\varepsilon
}\frac{\partial f_{s}\left(  \beta,\varepsilon,\zeta\right)  }{\partial
\varepsilon}.
\end{align}

In the limit of small amplitude $\left\vert \Psi\right\vert $, the gradient
term in the Hamiltonian (\ref{H}) with the coefficient $C$ is quadratic with
respect to $\left\vert \Psi\right\vert $, and the term with the coefficient
$E$ is quartic. So, in the vicinity of the critical temperature $T_{c}$ (where
the standard Ginzburg-Landau approach is applicable and $\left\vert
\Psi\right\vert $ is small), this quartic term becomes vanishingly small.
Hence it is absent in the standard GL theory, although far below $T_{c}$ it is
not negligible.

In order to study the time evolution of the ultracold Fermi gas, we need a
relation between real-time and Euclidean-time actions. The Euclidean-time
action $S\left(  \beta\right)  $ enters the partition function:%
\begin{equation}
\mathcal{Z}\propto\int\mathcal{D}\left[  \bar{\Psi},\Psi\right]  e^{-S\left(
\beta\right)  }, \label{Z}%
\end{equation}
while the real-time action $S\left(  t_{b},t_{a}\right)  $ enters the
transition amplitude:%
\begin{equation}
K\left(  t_{b},t_{a}\right)  =\int\mathcal{D}\left[  \bar{\Psi},\Psi\right]
e^{iS\left(  t_{b},t_{a}\right)  }. \label{K}%
\end{equation}
The correspondence between real-time and Euclidean-time actions is established
by the formal replacement in (\ref{Seucl}):%
\begin{equation}
\tau\rightarrow it\Leftrightarrow S\left(  \beta\right)  \rightarrow-iS\left(
t_{b},t_{a}\right)  . \label{corresp}%
\end{equation}
The real-time action can be then expressed as follows:%
\begin{equation}
S\left(  t_{b},t_{a}\right)  =\int_{t_{a}}^{t_{b}}dt\int d\mathbf{r}%
~\mathcal{L}, \label{S1}%
\end{equation}
where $\mathcal{L}$ is the field Lagrangian:%
\begin{equation}
\mathcal{L}=i\frac{D}{2}\left(  \bar{\Psi}\frac{\partial\Psi}{\partial
t}-\frac{\partial\bar{\Psi}}{\partial t}\Psi\right)  -\mathcal{H}. \label{L1}%
\end{equation}
Next, we use the regularized action, subtracting the background thermodynamic
potential $\Omega_{s}\left(  \left\vert \Psi_{\infty}\right\vert \right)  $
from the Hamiltonian. Here, $\left\vert \Psi_{\infty}\right\vert $ is the
modulus of the background order parameter. We express the order parameter
through the phase $\theta$ and amplitude $\left\vert \Psi\right\vert $ as
$\Psi=\left\vert \Psi\right\vert \exp\left(  i\theta\right)  $, using the
notation%
\begin{equation}
\left\vert \Psi\left(  \mathbf{r},t\right)  \right\vert =\left\vert
\Psi_{\infty}\right\vert \cdot a\left(  \mathbf{r},t\right)  \label{a}%
\end{equation}
where $a\left(  \mathbf{r},t\right)  \equiv\left\vert \Psi/\Psi_{\infty
}\right\vert $ is the amplitude modulation function. These notations are
suitable to describe localized disturbances such as a vortex or a soliton in
an otherwise homogeneous superfluid. For $r\rightarrow\infty,$ $a\left(
r\right)  \rightarrow a_{\infty}=1$. The procedure we follow consists in (1)
substituting the above form of $\Psi$ into the Lagrangian (\ref{L1}), and
interpreting the result as the Lagrangian for the amplitude and phase fields,
(2) extract from it the field equations for $a(\mathbf{r},t)$ and
$\theta(\mathbf{r},t),$ and (3) solve these equations.

First, we re-write the Lagrangian in the amplitude-phase representation:
\begin{equation}
\mathcal{L}=\int d\mathbf{r}\left(  -\kappa\left(  a\right)  a^{2}%
\frac{\partial\theta}{\partial t}-\mathcal{H}\right)  , \label{Lagr}%
\end{equation}
and the field Hamiltonian becomes%
\begin{equation}
\mathcal{H}=\int d\mathbf{r}\left[  \Omega_{s}\left(  a\right)  -\Omega
_{s}\left(  a_{\infty}\right)  +\frac{1}{2}\rho_{qp}\left(  a\right)  \left(
\nabla_{\mathbf{r}}a\right)  ^{2}+\frac{1}{2}\rho_{sf}\left(  a\right)
\left(  \nabla_{\mathbf{r}}\theta\right)  ^{2}\right]  . \label{H2}%
\end{equation}
The coefficient at the time derivative $\kappa$, the quantum pressure
coefficient $\rho_{qp}$, and the superfluid density $\rho_{sf}$ are determined
as follows:%
\begin{align}
\kappa\left(  a\right)   &  =D\left(  a\right)  \left\vert \Psi_{\infty
}\right\vert ^{2},\label{kappa}\\
\rho_{sf}\left(  a\right)   &  =\frac{C\left(  a\right)  }{m}\left\vert
\Psi\right\vert ^{2},\label{rhosf}\\
\rho_{qp}\left(  a\right)   &  =\frac{C\left(  a\right)  -4\left\vert
\Psi\right\vert ^{2}E\left(  a\right)  }{m}\left\vert \Psi_{\infty}\right\vert
^{2}. \label{rhoqp}%
\end{align}
They are, in general, depending on the amplitude $a$. It is easy to verify
analytically that the superfluid density given by (\ref{rhosf}) with (\ref{c})
is equivalent to the saddle-point superfluid density defined through a
\textquotedblleft phase twist\textquotedblright\ on the order parameter, as
phase gradients endow the pair condensate with a finite superfluid velocity
\cite{Taylor2006}. The pair condensate also resists gradients in the pair
density, $\nabla_{\mathbf{r}}a$, leading to a quantum pressure term as in
bosonic condensates.

The stationary wave and soliton solutions propagating with a constant velocity
$v_{S}$ obey the relation $f\left(  x,t\right)  =f\left(  x-v_{S}t\right)  $.
The stationary Lagrangian {\normalsize (\ref{Lagr})} is then given by:%
\begin{align}
\mathcal{L}  &  =\int_{-\infty}^{\infty}dx~\left\{  \kappa\left(  a\right)
a^{2}v_{S}\frac{\partial\theta}{\partial x}-\left[  \Omega_{s}\left(
a\right)  -\Omega_{s}\left(  a_{\infty}\right)  \right]  \right.  .\nonumber\\
&  \left.  -\frac{1}{2}\rho_{qp}\left(  a\right)  \left(  \frac{\partial
a}{\partial x}\right)  ^{2}-\frac{1}{2}\rho_{sf}\left(  a\right)  \left(
\frac{\partial\theta}{\partial x}\right)  ^{2}\right\}  \label{L}%
\end{align}
The macroscopic soliton dynamics is determined by the solutions of the
Lagrange equations for the Lagrangian (\ref{L}) for the phase and the
amplitude. The equation of motion for the phase reads:%
\begin{equation}
\frac{\partial}{\partial x}\left(  \rho_{sf}\left(  a\right)  \frac
{\partial\theta}{\partial x}-v_{S}\kappa a^{2}\right)  =0. \label{phase}%
\end{equation}
The general solution of this equation is%
\begin{equation}
\frac{\partial\theta}{\partial x}=\frac{1}{\rho_{sf}\left(  a\right)  }\left(
C+v_{S}\kappa\left(  a\right)  a^{2}\right)  \label{sol1}%
\end{equation}
with the integration constant $C$. Imposing the boundary condition
$\partial_{x}\theta\rightarrow0$ for $x\rightarrow\pm\infty$ corresponds to
the \textquotedblleft dark soliton\textquotedblright\ solution in which the
total change of phase accross the soliton is finite. This condition results in
the integration constant $C=-v_{S}\kappa_{\infty}$, where $\kappa_{\infty
}\equiv\kappa\left(  a_{\infty}\right)  $ with $a_{\infty}\equiv1$ is the bulk
value of the coefficient $\kappa\left(  a\right)  $. Hence the dark soliton
solution for the derivative $\partial_{x}\theta$ is%
\begin{equation}
\frac{\partial\theta}{\partial x}=\frac{v_{S}}{\rho_{sf}\left(  a\right)
}\left[  \kappa\left(  a\right)  a^{2}-\kappa_{\infty}\right]  \label{sol1a}%
\end{equation}
and the phase for the dark soliton can be determined explicitly:%
\begin{equation}
\theta\left(  x\right)  =v_{S}%
{\displaystyle\int\limits_{-\infty}^{x}}
\frac{\kappa\left(  a\left(  x^{\prime}\right)  \right)  a^{2}\left(
x^{\prime}\right)  -\kappa_{\infty}}{\rho_{sf}\left(  a\left(  x^{\prime
}\right)  \right)  }dx^{\prime}.
\end{equation}
The total phase change throughout the soliton is determined as the difference:%
\begin{equation}
\delta\theta\equiv\theta\left(  -\infty\right)  -\theta\left(  \infty\right)
\label{ch}%
\end{equation}
and results in the integral:%
\begin{equation}
\delta\theta=v_{S}%
{\displaystyle\int\limits_{-\infty}^{\infty}}
\frac{1}{\rho_{sf}\left(  a\left(  x\right)  \right)  }\left(  \frac
{\kappa_{\infty}}{a^{2}\left(  x\right)  }-\kappa\left(  a\left(  x\right)
\right)  \right)  dx. \label{Theta}%
\end{equation}

The Lagrange equation for the amplitude $a\left(  x\right)  $, with the
solution for the phase (\ref{sol1a}), takes the form%
\begin{equation}
\frac{\partial}{\partial x}\left(  \rho_{qp}\frac{\partial a}{\partial
x}\right)  =\frac{1}{2}\frac{\partial\rho_{qp}}{\partial a}\left(
\frac{\partial a}{\partial x}\right)  ^{2}+\frac{\partial\Omega_{s}}{\partial
a}-\frac{1}{2}v_{S}^{2}\frac{\partial}{\partial a}\left(  \frac{\left[
\kappa\left(  a\right)  a^{2}-\kappa_{\infty}\right]  ^{2}}{\rho_{sf}\left(
a\right)  }\right)  . \label{qampp1}%
\end{equation}
It also has the exact analytic solution. Imposing the boundary conditions%
\begin{equation}
\left.  \frac{\partial a\left(  x\right)  }{\partial x}\right\vert
_{x\rightarrow\pm\infty}=0,\quad\left.  a\left(  x\right)  \right\vert
_{x\rightarrow\pm\infty}=1, \label{BC}%
\end{equation}
and introducing the notations%
\begin{align}
X\left(  a\right)   &  \equiv\Omega_{s}\left(  a\right)  -\Omega_{s}\left(
a_{\infty}\right)  ,\label{X}\\
Y\left(  a\right)   &  \equiv\frac{\left[  \kappa\left(  a\right)
a^{2}-\kappa_{\infty}\right]  ^{2}}{2\rho_{sf}\left(  a\right)  }, \label{Y}%
\end{align}
we arrive at the symmetric solution for the coordinate $x$ as a function of
the relative amplitude $a$:%
\begin{equation}
x=\pm\frac{1}{\sqrt{2}}%
{\displaystyle\int\limits_{a_{0}}^{a}}
\frac{\sqrt{\rho_{qp}\left(  a^{\prime}\right)  }}{\sqrt{X\left(  a^{\prime
}\right)  -v_{S}^{2}Y\left(  a^{\prime}\right)  }}da^{\prime}. \label{xres}%
\end{equation}
The amplitude modulation function at the soliton center, $a_{0}\equiv a\left(
x=0\right)  $, is determined by the equation%
\begin{equation}
X\left(  a_{0}\right)  -v_{S}^{2}Y\left(  a_{0}\right)  =0. \label{a0}%
\end{equation}

The exact analytic solutions obtained above for a dark soliton allow us to
consider the soliton dynamics in terms of the macroscopic integrals of motion
for the soliton: the momentum and the energy. They are determined in the next
section through the canonical definitions of the classical Hamilton dynamics
in the spirit of Ref. \cite{SolReview}.

\section{Integrals of motion \label{sec:rels}}

The total soliton momentum is obtained by differentiating the Lagrangian with
respect to the soliton velocity:%
\begin{equation}
\mathcal{P}_{S}^{\left(  tot\right)  }\equiv\frac{\partial\mathcal{L}%
}{\partial v_{S}}. \label{def}%
\end{equation}
With the Lagrangian (\ref{L}), the total soliton momentum is%
\begin{equation}
\mathcal{P}_{S}^{\left(  tot\right)  }=\int_{-\infty}^{\infty}dx~\kappa\left(
a\right)  a^{2}\frac{\partial\theta}{\partial x}. \label{Ptot1}%
\end{equation}
Using the exact solution for the phase, we express the total momentum
explicitly:
\begin{equation}
\mathcal{P}_{S}^{\left(  tot\right)  }\left(  v_{S}\right)  =v_{S}%
\int_{-\infty}^{\infty}dx~\frac{\kappa\left(  a\right)  }{\rho_{sf}\left(
a\right)  }\left(  \kappa\left(  a\right)  -\frac{\kappa_{\infty}}{a^{2}%
}\right)  . \label{Ps}%
\end{equation}

As established in Ref. \cite{SolReview}, the total momentum $\mathcal{P}%
_{S}^{\left(  tot\right)  }$ refers to both the soliton and the uniform
background. Therefore, in order to obtain the \textquotedblleft
pure\textquotedblright\ soliton momentum, the background contribution must be
subtracted. As long as we follow the scheme of Ref. \cite{SolReview}, the
background part of the momentum $\mathcal{P}_{S}^{\left(  u\right)  }$ is
obtained as%
\begin{equation}
\mathcal{P}_{S}^{\left(  u\right)  }\left(  v_{S}\right)  =-\kappa_{\infty
}\left[  \delta\theta\left(  v_{S}\right)  -\delta\theta\left(  0\right)
\right]  , \label{Pu}%
\end{equation}
where $\kappa_{\infty}\equiv$ $\kappa\left(  a_{\infty}\right)  $ with
$a_{\infty}\equiv1$, $\delta\theta\left(  v_{S}\right)  $ is the total change
of the phase determined by (\ref{ch}) and (\ref{Theta}), which depends on the
soliton velocity, and $\delta\theta\left(  0\right)  =\pi$. Performing the
subtraction of the background contribution, we arrive at the result%
\begin{equation}
\mathcal{P}_{S}\left(  v_{S}\right)  =v_{S}\int_{-\infty}^{\infty}%
dx~\frac{\left[  \kappa\left(  a\right)  a^{2}-\kappa_{\infty}\right]  ^{2}%
}{\rho_{sf}\left(  a\right)  }-\pi\kappa_{\infty}. \label{PS0}%
\end{equation}
We use the replacement of the integration variable and solution for the
amplitude (\ref{xres}). The resulting soliton momentum is then given by:%
\begin{equation}
\mathcal{P}_{S}\left(  v_{S}\right)  =2\sqrt{2}v_{S}\int_{a_{0}}^{1}%
\frac{\sqrt{\rho_{qp}\left(  a\right)  }Y\left(  a\right)  }{\sqrt{X\left(
a\right)  -v_{S}^{2}Y\left(  a\right)  }}da-\pi\kappa_{\infty}. \label{PS1}%
\end{equation}
It is easy to check that the subtraction of $\pi\kappa_{\infty}$ in
(\ref{PS1}) is necessary in order to ensure the zero-velocity limit
$\lim_{v_{S}\rightarrow0}\mathcal{P}_{S}\left(  v_{S}\right)  =0$.

The soliton energy is defined accordingly to the rule of the classical
mechanics:%
\begin{equation}
\mathcal{E}_{S}\equiv v_{S}\mathcal{P}_{S}-\mathcal{L}=\mathcal{H}.
\end{equation}
Using the exact solutions for the amplitude and the phase, we arrive at a
simple expression for the energy:%
\begin{equation}
\mathcal{E}_{S}\left(  v_{S}\right)  =2\int_{-\infty}^{\infty}dx~\left[
\Omega_{s}\left(  a\right)  -\Omega_{s}\left(  a_{\infty}\right)  \right]  .
\label{ES}%
\end{equation}
As for the momentum, the replacement of variables ($a$ instead of $x$) yields
the soliton energy expressed through the integral over the amplitude
modulation:%
\begin{equation}
\mathcal{E}_{S}\left(  v_{S}\right)  =2\sqrt{2}\int_{a_{0}}^{1}\frac
{\sqrt{\rho_{qp}\left(  a\right)  }X\left(  a\right)  }{\sqrt{X\left(
a\right)  -v_{S}^{2}Y\left(  a\right)  }}da. \label{E2}%
\end{equation}

Next, we check whether the relation (\ref{dynam}) is fulfilled for a dark
soliton within the present GL-like approach. The derivative $\partial
\mathcal{E}_{S}/\partial\mathcal{P}_{S}$ can be expressed as%
\begin{equation}
\frac{\partial\mathcal{E}_{S}}{\partial\mathcal{P}_{S}}=\left(  \frac
{\partial\mathcal{P}_{S}}{\partial v_{S}}\right)  ^{-1}\left(  \frac
{\partial\mathcal{E}_{S}}{\partial v_{S}}\right)  . \label{id}%
\end{equation}
The details of the calculation for the derivatives are represented in the
Appendix A. Here, we represent the final results:%
\begin{align}
\frac{\partial\mathcal{E}_{S}}{\partial v_{S}}  &  =2\sqrt{2}v_{S}\int_{a_{0}%
}^{1}\left(  \frac{\sqrt{\rho_{qp}\left(  a\right)  }X\left(  a\right)
Y\left(  a\right)  }{\left[  X\left(  a\right)  -Y\left(  a\right)  v_{S}%
^{2}\right]  ^{3/2}}-\frac{\sqrt{\rho_{qp}\left(  a_{0}\right)  }X\left(
a_{0}\right)  Y\left(  a_{0}\right)  }{\left[  G\left(  a_{0}\right)  \right]
^{3/2}\left(  a-a_{0}\right)  ^{3/2}}\right)  da\nonumber\\
&  -4\sqrt{2}v_{S}\frac{\sqrt{\rho_{qp}\left(  a_{0}\right)  }X\left(
a_{0}\right)  Y\left(  a_{0}\right)  }{\left[  G\left(  a_{0}\right)  \right]
^{3/2}\sqrt{1-a_{0}}}, \label{dE1}%
\end{align}
and%
\begin{align}
\frac{\partial\mathcal{P}_{S}}{\partial v_{S}}  &  =2\sqrt{2}\int_{a_{0}}%
^{1}\left(  \frac{\sqrt{\rho_{qp}\left(  a\right)  }X\left(  a\right)
Y\left(  a\right)  }{\left[  X\left(  a\right)  -Y\left(  a\right)  v_{S}%
^{2}\right]  ^{3/2}}-\frac{\sqrt{\rho_{qp}\left(  a_{0}\right)  }X\left(
a_{0}\right)  Y\left(  a_{0}\right)  }{\left[  G\left(  a_{0}\right)  \right]
^{3/2}\left(  a-a_{0}\right)  ^{3/2}}\right)  da\nonumber\\
&  -4\sqrt{2}\frac{\sqrt{\rho_{qp}\left(  a_{0}\right)  }X\left(
a_{0}\right)  Y\left(  a_{0}\right)  }{\left[  G\left(  a_{0}\right)  \right]
^{3/2}\sqrt{1-a_{0}}}. \label{dP1}%
\end{align}
where the function $G\left(  a\right)  $ is given by formula (\ref{G}).

The effective mass of the dark soliton can be introduced as in Ref.
\cite{SolReview}:%
\begin{equation}
M_{S}\equiv\frac{\partial\mathcal{P}_{S}}{\partial v_{S}}. \label{Mass1}%
\end{equation}
Therefore formula (\ref{dP1}) allows us to determine the effective mass of the
dark soliton explicitly:%
\begin{align}
M_{S}\left(  v_{S}\right)   &  =2\sqrt{2}\int_{a_{0}}^{1}\left(  \frac
{\sqrt{\rho_{qp}\left(  a\right)  }X\left(  a\right)  Y\left(  a\right)
}{\left[  X\left(  a\right)  -Y\left(  a\right)  v_{S}^{2}\right]  ^{3/2}%
}-\frac{\sqrt{\rho_{qp}\left(  a_{0}\right)  }X\left(  a_{0}\right)  Y\left(
a_{0}\right)  }{\left[  G\left(  a_{0}\right)  \right]  ^{3/2}\left(
a-a_{0}\right)  ^{3/2}}\right)  da\nonumber\\
&  +\frac{4\sqrt{2}\sqrt{\rho_{qp}\left(  a_{0}\right)  }X\left(
a_{0}\right)  Y\left(  a_{0}\right)  }{\left[  G\left(  a_{0}\right)  \right]
^{3/2}\sqrt{1-a_{0}}}. \label{Mass}%
\end{align}

When comparing (\ref{de}) with (\ref{dp1}) we find the exact analytic
relation:%
\begin{equation}
\frac{\partial\mathcal{E}_{S}}{\partial v_{S}}=v_{S}\frac{\partial
\mathcal{P}_{S}}{\partial v_{S}}. \label{rel}%
\end{equation}
The identity (\ref{id}) combined with (\ref{rel}) gives us the same equation
as in Ref. \cite{SolReview}:%
\begin{equation}
\frac{\partial\mathcal{E}_{S}}{\partial\mathcal{P}_{S}}=v_{S}. \label{dynam}%
\end{equation}
This equation shows that the soliton described within the present formalism
obeys the classic Hamilton dynamics, i. e., moves like a particle. This
behavior holds even in the most general case -- for arbitrary $\Omega
_{s}\left(  a\right)  $ and amplitude-dependent coefficients in the
{\normalsize effective field action}.

\section{Results and discussion \label{sec:results}}

The subsequent numerical analysis is restricted to the case of a balanced
Fermi gas, where the populations of the \textquotedblleft
spin-up\textquotedblright\ and \textquotedblleft spin-down\textquotedblright%
\ fermions are equal. The soliton parameters are calculated here in two
approximations: (1) within the saddle-point approximation for $\left\vert
\Psi_{\infty}\right\vert $ and the chemical potential $\mu,$ which are
obtained using mean-field number and gap equations, and (2) accounting for
fluctuations about the saddle point. The Gaussian fluctuations are included
here within the same scheme as in Refs. \cite{deMelo1993,PRA2008}, through the
renormalization of the chemical potential of the Fermi gas. In both cases, the
results obtained within our effective field theory are compared with results
obtained with Bogoliubov-de Gennes theory applied at unitarity, from Ref.
\cite{Liao2011}.

It should be noted that the effective action for the pair field (\ref{Seucl})
has been derived in Ref. \cite{ExtGL2} using a gradient expansion up to second
order in spatial gradients and in imaginary time gradients. This is consistent
with the assumption that the pair field slowly varies in space and time.
Consequently, keeping the coordinate- and time dependence of the coefficients
$C,E$ that appear in front of the second-order gradient factors is, strictly
speaking, beyond the second-order and may lead to artefacts in the limiting
case when simultaneously $T\rightarrow0$ and $a\rightarrow0$.
{\normalsize Therefore we} keep the coefficients $C\left(  a\right)  $ and
$E\left(  a\right)  $ in the present numerical analysis equal to their
background (bulk) values $C\left(  a_{\infty}\right)  $ and $E\left(
a_{\infty}\right)  $. On the contrary, in the thermodynamic potential
$\Omega_{s}\left(  a\right)  $ and in the first-order terms of the gradient
expansion in the effective action (time derivatives) we keep the amplitude
dependence of the coefficients.

The results are presented in Figs. 1-8. Each figure -- except Fig.3 -- shows
how a solitonic property depends on the soliton velocity $v_{S}$, and is
divided in six panels. The top row contains results for the BCS regime
($a_{s}=-0.5$), the middle row for unitarity ($a_{s}=0$) and the bottom row
for the BEC regime ($a_{s}=1$). The left column shows the results using the
mean-field value for $\left\vert \Psi_{\infty}\right\vert ,\mu$, and the right
column shows the results including fluctuations in $\left\vert \Psi_{\infty
}\right\vert ,\mu$. Dots in the figures represent Bogoliubov-de Gennes results
from Ref. \cite{Liao2011}.

Fig. 1 shows the ratio of the amplitude of the order parameter $\Psi$ at the
soliton center to the bulk value $\left\vert \Psi_{\infty}\right\vert $ (in
other words, the amplitude modulation function $\left.  a\left(  x\right)
\right\vert _{x=0}\equiv a_{0}$) as a function of the soliton velocity $v_{S}%
$. The dependence $a_{0}\left(  v_{S}\right)  $ is close to a linear function
for all considered temperatures and scattering lengths. The slope of that
linear dependence rises with increasing temperature. The critical velocity
$v_{S}^{\left(  c\right)  }$, when $a_{0}=1$, indicates a breakdown of the
soliton state: a soliton does not exist for $v_{S}>v_{S}^{\left(  c\right)  }%
$. The critical velocity obtained in the present work is close to the sound
velocity determined in Ref. \cite{Liao2011} for the unitarity regime as
$c=v_{F}\sqrt{\mu/(3E_{F})}$, where $v_{F}$ is the Fermi velocity (equal to
$v_{F}=2$ in the present units). The critical velocity diminishes when
temperature rises. The obtained close-to-linear dependence of $a_{0}\left(
v_{S}\right)  $ is in line with the results of the BdG theory from Ref.
\cite{Liao2011}, but increases slightly more slowly than the BdG solution in
the unitarity regime.

In Fig. 2, we plot the relative fermion density dip at the soliton center
$n_{0}/n_{\infty}$ (where $n_{\infty}$ is the bulk fermion density) as a
function of the soliton velocity for the same initial parameters as in Fig. 1.
It is clear that even though the pair density at the soliton center may become
small (as $v_{S}\rightarrow0$), the soliton partially fills up with unpaired
atoms, leading to $n_{0}/n_{\infty}>a_{0}$. The fermion density is determined
here in two ways: (1) within the mean-field local density approximation (LDA),
using the formula%
\begin{equation}
n^{\left(  \text{LDA}\right)  }=-\frac{\partial\Omega_{s}}{\partial\mu
},\label{na}%
\end{equation}
and (2) accounting for the \textquotedblleft gradient part\textquotedblright%
\ of the density -- provided by the gradient terms in the Hamiltonian
(\ref{H2}):%
\begin{equation}
n^{\left(  \operatorname{grad}\right)  }=-\frac{\partial\Omega_{s}}%
{\partial\mu}-\frac{1}{2}\frac{\partial\rho_{qp}}{\partial\mu}\left(  \nabla
a\right)  ^{2}-\frac{1}{2}\frac{\partial\rho_{sf}}{\partial\mu}\left(
\nabla\theta\right)  ^{2}.\label{nb}%
\end{equation}
When comparing to each other the graphs for $n_{0}/n_{\infty}$ with different
scattering lengths, we see that the relative contribution of the gradient part
of the density is more significant for weaker coupling strengths and for lower
temperatures: the highest difference between $n^{\left(  \operatorname{grad}%
\right)  }$ and $n^{\left(  \text{LDA}\right)  }$ occurs in the BCS regime at
the lowest considered temperature. The relative depth of the fermion density
dip qualitatively follows the BdG results, being slightly smaller at $v_{S}%
=0$, and showing a less expressed dependence on the soliton velocity with
respect to the BdG data. Fig. 3 shows the overall density profile of the
soliton, as a function of distance from the center of the soliton. Also this
is seen to reproduce the BdG results well. However, in the BCS regime, the
Friedel oscillations of the density obtained in the BdG calculations do not
agree with the results in the current formalism, even though the profile
closer to $x=0$ still is the same in both formalisms. The disagreement between
BdG and EFT results in the BCS regime can be due to the fact that higher-order
terms of the gradient expansion could play a more prominent role at weak coupling.

Fig. 4 shows the total phase difference $\delta\theta$ through the soliton.
The comparison with the BdG results is performed for unitarity regime. As for
the density, the BdG calculation gives a faster decrease of the total phase
change as a function of $v_{S}$ with respect to that calculated using EFT. As
stated in Ref. \cite{Liao2011}, one of the key results of the BdG approach
applied to dark solitons is a drastic qualitative difference of the phase
difference with respect to that obtained using the Gross-Pitaevskii (GP)
equation. The GP method yields $\cos\left[  \delta\theta\left(  v_{S}\right)
\right]  \propto v_{S}$, so that at small velocities one obtains $\pi
-\delta\theta\left(  v_{S}\right)  \propto v_{S}^{2}$. On the contrary, the
BdG approach results in a linear dependence $\delta\theta\left(  v_{S}\right)
$, at least at small $v_{S}$. Within the effective field theory, $\delta
\theta\left(  v_{S}\right)  $ is linear at small velocities, that is more
realistic with respect to the GP results and closer to the BdG data. However,
a quantitative difference between the present results and BdG at unitarity remains.

In Fig. 5, the soliton energy $E_{S}\left(  v_{S}\right)  $ is plotted and
compared with the BdG data for all three regimes (BCS, unitarity and BEC). The
best coincidence of the soliton energy calculated within the BDG and EFT
methods is obtained in the BEC regime. at weaker couplings, the maximum of the
energy (at $v_{S}=0$) provided by the BdG method is higher than that obtained
within EFT. At small velocities, the energy is approximately quadratic with a
negative second derivative, indicating a negative effective mass of the
soliton. The soliton energy falls down to zero when the velocity reaches its
critical value $v_{S}^{\left(  c\right)  }$. Fig. 6 shows the soliton momentum
$P_{S}\left(  v_{S}\right)  $. At small $v_{S}$, the dependence $P_{S}\left(
v_{S}\right)  $ is close to a linear function with a negative slope (which
also indicates a negative effective mass of the soliton). With increasing
velocity, the soliton momentum ends at a finite value when $v_{S}$ reaches a
critical value $v_{S}^{\left(  c\right)  }$ (different for different coupling
strengths and temperatures). The energy-momentum relation (\ref{dynam}) has
been numerically checked in the present calculation for the data represented
in Figs. 5 and 6. This verification has shown that it is indeed fulfilled.

In Fig. 7, we plot the number of fermions in the soliton cloud $N_{S}$ (per
unit area in the $yz$-plane), determined by the integral%
\begin{equation}
N_{S}=\int_{-\infty}^{\infty}\left[  n\left(  x\right)  -n\left(
\infty\right)  \right]  dx. \label{NS}%
\end{equation}
When multiplied by the fermion mass $m$ (here, $m=1/2$ as in our previous
calculations), the fermion number yields the \textquotedblleft physical
mass\textquotedblright\ of a soliton $mN_{S}$ \cite{Liao2011}. The fermion
density in a dark soliton is lower than the bulk fermion density. Hence the
number of fermions $N_{S}$, as well as the physical mass of a dark soliton, is
negative. The absolute number of fermions $\left\vert N_{S}\right\vert $ in a
soliton monotonously decreases as a function of the velocity $v_{S}$. The
absolute value $\left\vert N_{S}\right\vert $ gradually rises with an
increasing coupling strength and diminishes with an increasing temperature.
The comparison with the BdG data \cite{Liao2011} is possible at present for
the unitarity regime. At small velocities, the BdG and EFT results for the
number of fermions in the soliton match each other very well. The increase of
the number of fermions in the soliton as a function of $v_{S}$ is, however,
faster for the BdG method than for the EFT.

Finally, Fig. 8 represents the effective mass of the soliton determined by
formulae (\ref{Mass1}), (\ref{Mass}). As follows from (\ref{dynam}), the other
definition of the effective mass (e. g., in Ref. \cite{Liao2011}),%
\begin{equation}
M_{S}=\frac{1}{v_{S}}\frac{\partial\mathcal{E}_{S}}{\partial v_{S}}
\label{mass1}%
\end{equation}
is equivalent to (\ref{Mass}). The effective and physical masses of a dark
soliton are, in general, different \cite{Scott}. The effective mass in the BCS
and unitarity regimes non-monotonously behaves as a function of velocity. In
the BEC regime, however, $\left\vert M_{S}\right\vert $ monotonously
{\normalsize decreases} when $v_{S}$ rises. At unitarity and at weak coupling,
the effective and physical masses are rather close to each other. In the BEC
regime the effective mass of the soliton appears to be larger (in absolute
value) than the physical mass.

The comparison of the soliton parameters calculated with and without effect of
the Gaussian fluctuations shows that the range of the soliton velocities
$v_{S}<v_{S}^{\left(  c\right)  }$ where the soliton exists is relatively
slightly influenced by the fluctuations in all three regimes: the BCS regime
($1/a_{s}=-0.5$), at unitarity ($1/a_{s}=0$), and in the BEC regime
($1/a_{s}=1$). {\normalsize The} behavior of the amplitude modulation
function, the fermion density and the phase difference exhibit the same trend.
On the contrary, the relative change of the soliton energy, the soliton
momentum and both physical and effective masses is re-scaled {\normalsize more
strongly}: even in the BCS regime it is not small. Qualitatively, the
dependence of the calculated dark soliton parameters on the soliton velocity,
the temperature and the coupling strength is similar to that obtained within
the mean-field approach. The effect of the fluctuations consists in a scaling
of the calculated parameters due to the renormalization of the
{\normalsize density. }The critical temperature for a Fermi gas obtained
accounting for fluctuations is lower than the mean-field critical temperature,
especially at sufficiently strong coupling (at unitarity and in the BEC
regime, where $T_{c}$ with fluctuations tends to a constant value when
$1/a_{s}\rightarrow+\infty$, contrary to the mean-field critical temperature).
Consequently, the dark soliton parameters in the BEC regime calculated
accounting for fluctuations are significantly more sensitive to temperature
than those calculated within the mean-field approximation.

\section{Conclusions \label{sec:conclusions}}

Within the effective field theory, we have derived the analytic solution of
the field equations which describes a dark soliton in a superfluid Fermi gas
with $s$-wave pairing for arbitrary temperatures below $T_{c}$ and for
arbitrary values of the inverse scattering length, encompassing the BCS-BEC
crossover regime. The macroscopic parameters of the dark soliton (modulus of
the order parameter, and phase profile) are analytically expressed by assuming
that the order parameter has the usual solitonic $f(x-v_{s}t)$ time
dependence, expressing the conservation of the soliton shape as it travels at
constant speed. This assumption does not allow to investigate the decay of the
soliton due to the snake instability
\cite{CetoliPRA88,BulgacPRL112,WenPRA88,ToikkaPRA87}, and does not take into
account the inhomogeneity or anisotropy due to trapping, even though the
effective field theory itself allows in principle to investigate this decay
dynamics as well. Here, we compute the density profile, and with it the
filling of the soliton \textquotedblleft core\textquotedblright\ by unpaired
fermions. The exact energy-momentum relation for the soliton has been derived,
showing that the soliton as a whole obeys classical Hamiltonian dynamics, with
a well-determined effective mass, depending of the velocity. The comparison of
the soliton parameters obtained within the mean-field approach with the
results of the calculation using the BdG equations have shown that in the BEC
regime, the {\normalsize EFT} provides an excellent agreement with BdG for all
temperatures below $T_{c}$. In the BCS regime and at unitarity, the obtained
analytic solutions match well the numeric BdG results for sufficiently high
temperatures, nevertheless well below $T_{c}$. That indicates a substantial
extension of the range of validity of the present method with respect to the
standard GL.

Besides the mean-field calculation, we have taken into account the Gaussian
fluctuations through the renormalization of the chemical potential of the
fermions. This renormalization keeps the qualitative picture of the soliton
similar to that obtained within the mean-field approximation, but leads to
quantitative changes of the soliton parameters and distributions.

\begin{acknowledgments}
We are grateful to J. Brand, L. Salasnic and G.C.\ Strinati for valuable
discussions. This work was supported by FWO-V projects G.0370.09N, G.0180.09N,
G.0115.12N, G.0119.12N, the WOG WO.033.09N (Belgium).
\end{acknowledgments}

\appendix

\section{Derivatives of the energy and the momentum \label{sec:App}}

When we take the derivative $\partial\mathcal{E}_{S}/\partial v_{S}$ in
(\ref{E2}) straightforwardly, the integral over $a$ becomes divergent at
$a\rightarrow a_{0}$ due to the appearance of the factor $\left[  X\left(
a\right)  -v_{S}^{2}Y\left(  a\right)  \right]  ^{3/2}$ in the denominator. To
remove divergencies, we consider the auxiliary expression with the parameter
$\delta>0$%
\begin{equation}
\mathcal{E}_{S}\left(  v_{S},\delta\right)  \equiv2\sqrt{2}\int_{a_{0}}%
^{1}\frac{\sqrt{\rho_{qp}\left(  a\right)  }X\left(  a\right)  }%
{\sqrt{X\left(  a\right)  -v_{S}^{2}Y\left(  a\right)  +\delta}}da.
\label{Es3}%
\end{equation}
The function $\mathcal{E}_{S}\left(  v_{S},\delta\right)  $ turns to
$\mathcal{E}_{S}\left(  v_{S}\right)  $ in the limit $\delta\rightarrow+0$.

Differentiating $\mathcal{E}_{S}\left(  v_{S},\delta\right)  $ with respect to
$v_{S}$ we obtain the result:%
\begin{equation}
\frac{\partial\mathcal{E}_{S}\left(  v_{S},\delta\right)  }{\partial v_{S}%
}=2\sqrt{2}v_{S}\int_{a_{0}}^{1}\frac{\sqrt{\rho_{qp}\left(  a\right)
}X\left(  a\right)  Y\left(  a\right)  }{\left[  X\left(  a\right)
+\delta-Y\left(  a\right)  v_{S}^{2}\right]  ^{3/2}}da-\frac{\partial a_{0}%
}{\partial v_{S}}\frac{2\sqrt{2}}{\sqrt{\delta}}\sqrt{\rho_{qp}\left(
a_{0}\right)  }X\left(  a_{0}\right)  . \label{dE}%
\end{equation}
The regularization of (\ref{dE}) is found using the Taylor series of the
denominator about the point $a=a_{0}\,$. Accounting for (\ref{a0}), we find
the expansion of the function $X\left(  a\right)  -v_{S}^{2}Y\left(  a\right)
+\delta$ about $a_{0}$:%
\begin{equation}
X\left(  a\right)  -v_{S}^{2}Y\left(  a\right)  +\delta=G\left(  a_{0}\right)
\left(  a-a_{0}\right)  +\delta+\ldots\label{Taylor}%
\end{equation}
where we have denoted the function%
\begin{equation}
G\left(  a\right)  \equiv\frac{\partial X\left(  a\right)  }{\partial a}%
-v_{S}^{2}\frac{\partial Y\left(  a\right)  }{\partial a}. \label{G}%
\end{equation}
Next, consider the auxiliary integral%
\[
\int_{a_{0}}^{1}\frac{1}{\left(  G\cdot\left(  a-a_{0}\right)  +\delta\right)
^{3/2}}da=2\frac{-\sqrt{\delta}+\sqrt{G-Ga_{0}+\delta}}{\sqrt{G-Ga_{0}+\delta
}G\sqrt{\delta}}.
\]
In the limit of small $\delta$, this integral behaves as $1/\sqrt{\delta}$:%
\[
\int_{a_{0}}^{1}\frac{1}{\left(  G\cdot\left(  a-a_{0}\right)  +\delta\right)
^{3/2}}da=\frac{2}{\sqrt{\delta}G}-\frac{2}{G^{3/2}\sqrt{1-a_{0}}}+O\left(
\delta\right)  .
\]
Thus we can express the factor $2/(G\sqrt{\delta})$ as%
\begin{equation}
\frac{2}{G\sqrt{\delta}}=\int_{a_{0}}^{1}\frac{1}{\left(  G\cdot\left(
a-a_{0}\right)  +\delta\right)  ^{3/2}}da+\frac{2}{G^{3/2}\sqrt{1-a_{0}}%
}+O\left(  \delta\right)  . \label{fact}%
\end{equation}
The factor $\partial a_{0}/(\partial v_{S})$ is determined as follows:%
\begin{equation}
\frac{\partial a_{0}}{\partial v_{S}}=-\frac{\frac{\partial\left(  X\left(
a_{0}\right)  -Y\left(  a_{0}\right)  v_{S}^{2}\right)  }{\partial v_{S}}%
}{\frac{\partial\left(  X\left(  a_{0}\right)  -Y\left(  a_{0}\right)
v_{S}^{2}\right)  }{\partial a_{0}}}=2v_{S}\frac{Y\left(  a_{0}\right)
}{G\left(  a_{0}\right)  }.
\end{equation}
Using these results the derivative $\partial\mathcal{E}_{S}\left(
\delta\right)  /\partial v_{S}$ is transformed to the expression:%
\begin{align}
\frac{\partial\mathcal{E}_{S}\left(  \delta\right)  }{\partial v_{S}}  &
=2\sqrt{2}v_{S}\int_{a_{0}}^{1}\left(  \frac{\sqrt{\rho_{qp}\left(  a\right)
}X\left(  a\right)  Y\left(  a\right)  }{\left[  X\left(  a\right)  -v_{S}%
^{2}Y\left(  a\right)  +\delta\right]  ^{3/2}}-\frac{\sqrt{\rho_{qp}\left(
a_{0}\right)  }X\left(  a_{0}\right)  Y\left(  a_{0}\right)  }{\left[
G\left(  a_{0}\right)  \left(  a-a_{0}\right)  +\delta\right]  ^{3/2}}\right)
da\nonumber\\
&  -4\sqrt{2}v_{S}\frac{\sqrt{\rho_{qp}\left(  a_{0}\right)  }X\left(
a_{0}\right)  Y\left(  a_{0}\right)  }{\left[  G\left(  a_{0}\right)  \right]
^{3/2}\sqrt{1-a_{0}}}. \label{dE3}%
\end{align}
The resulting integral over $a$ converges. Hence we can explicitly set
$\delta\rightarrow0$ in (\ref{dE3}), yielding the regularized expression for
the derivative $\partial\mathcal{E}_{S}/\partial v_{S}$:%
\begin{align}
\frac{\partial\mathcal{E}_{S}}{\partial v_{S}}  &  =2\sqrt{2}v_{S}\int_{a_{0}%
}^{1}\left(  \frac{\sqrt{\rho_{qp}\left(  a\right)  }X\left(  a\right)
Y\left(  a\right)  }{\left[  X\left(  a\right)  -Y\left(  a\right)  v_{S}%
^{2}\right]  ^{3/2}}-\frac{\sqrt{\rho_{qp}\left(  a_{0}\right)  }X\left(
a_{0}\right)  Y\left(  a_{0}\right)  }{\left[  G\left(  a_{0}\right)  \right]
^{3/2}\left(  a-a_{0}\right)  ^{3/2}}\right)  da\nonumber\\
&  -4\sqrt{2}v_{S}\frac{\sqrt{\rho_{qp}\left(  a_{0}\right)  }X\left(
a_{0}\right)  Y\left(  a_{0}\right)  }{\left[  G\left(  a_{0}\right)  \right]
^{3/2}\sqrt{1-a_{0}}}. \label{de}%
\end{align}
Repeating the same steps for the derivative of the momentum, we arrive at the
regularized expression:%
\begin{align}
\frac{\partial\mathcal{P}_{S}}{\partial v_{S}}  &  =2\sqrt{2}\int_{a_{0}}%
^{1}\left(  \frac{\sqrt{\rho_{qp}\left(  a\right)  }X\left(  a\right)
Y\left(  a\right)  }{\left[  X\left(  a\right)  -Y\left(  a\right)  v_{S}%
^{2}\right]  ^{3/2}}-\frac{\sqrt{\rho_{qp}\left(  a_{0}\right)  }X\left(
a_{0}\right)  Y\left(  a_{0}\right)  }{\left[  G\left(  a_{0}\right)  \right]
^{3/2}\left(  a-a_{0}\right)  ^{3/2}}\right)  da\nonumber\\
&  -4\sqrt{2}\frac{\sqrt{\rho_{qp}\left(  a_{0}\right)  }X\left(
a_{0}\right)  Y\left(  a_{0}\right)  }{\left[  G\left(  a_{0}\right)  \right]
^{3/2}\sqrt{1-a_{0}}}. \label{dp1}%
\end{align}

\newpage

\textbf{\textsl{FIGURES}}

\bigskip%

\begin{figure}
[h]
\begin{center}
\includegraphics[
height=6.5138in,
width=5.1059in
]%
{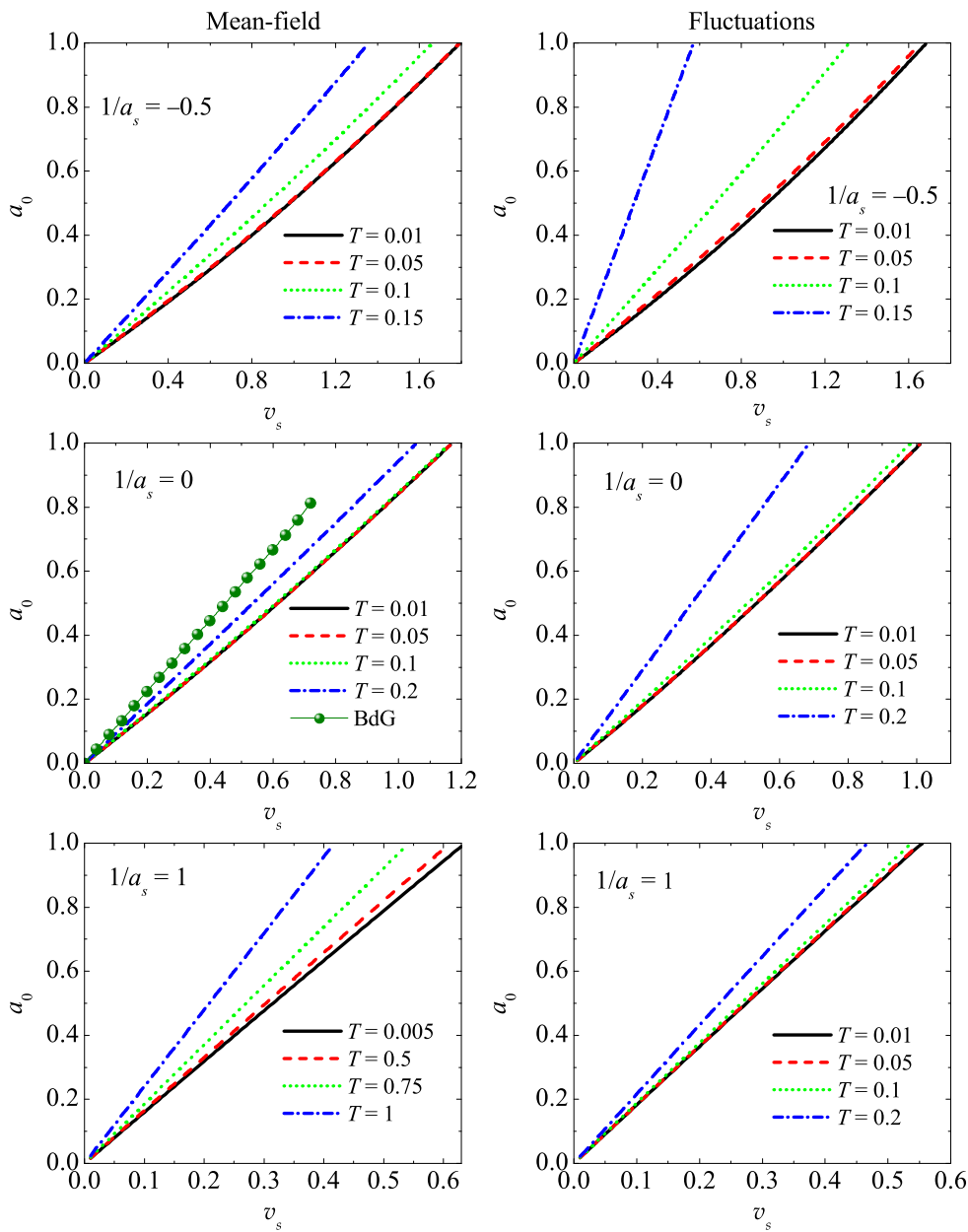}%
\caption{Amplitude modulation function at the soliton center depending on the
soliton velocity $v_{S}$ for different scattering lengths and temperatures.
\emph{Left-hand panels}: the mean-field calculation. \emph{Right-hand panels}:
the results obtained accounting for Gaussian fluctuations. The symbols (full
dots) show the results of the BdG theory from Ref. \cite{Liao2011}.}%
\end{center}
\end{figure}

\newpage%

\begin{figure}
[h]
\begin{center}
\includegraphics[
height=6.429in,
width=5.1327in
]%
{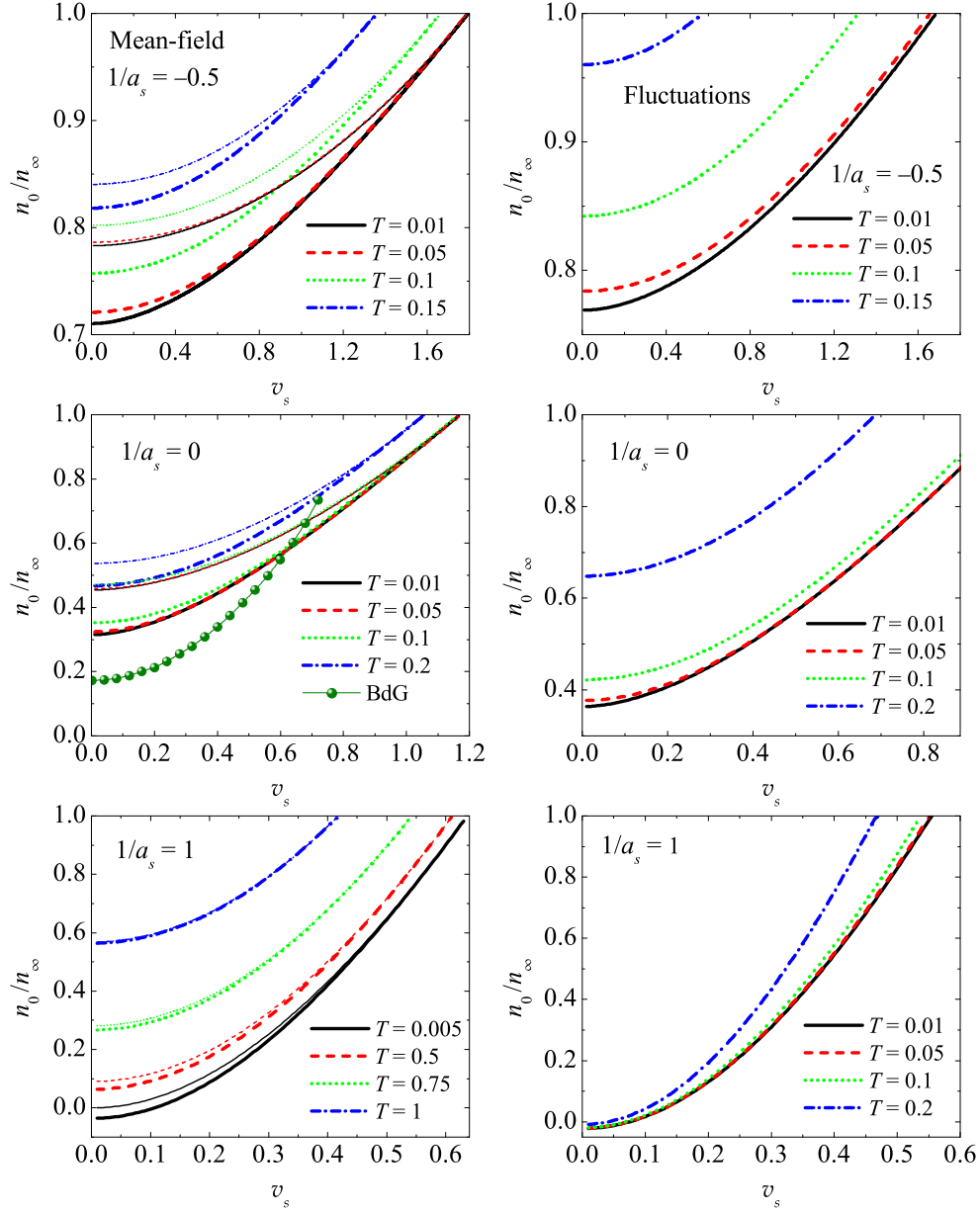}%
\caption{Relative fermion density at the soliton center $n_{0}/n_{\infty}$ as
a function on the soliton velocity $v_{S}$ for different scattering lengths
and temperatures. \emph{Heavy curves}: the density calculated accounting for
gradient terms. \emph{Thin curves}: the density calculated within LDA. The
full dots show the results of the BdG theory from Ref. \cite{Liao2011}.}%
\end{center}
\end{figure}

\newpage%

\begin{figure}
[h]
\begin{center}
\includegraphics[
height=3.8in,
width=4.804in
]%
{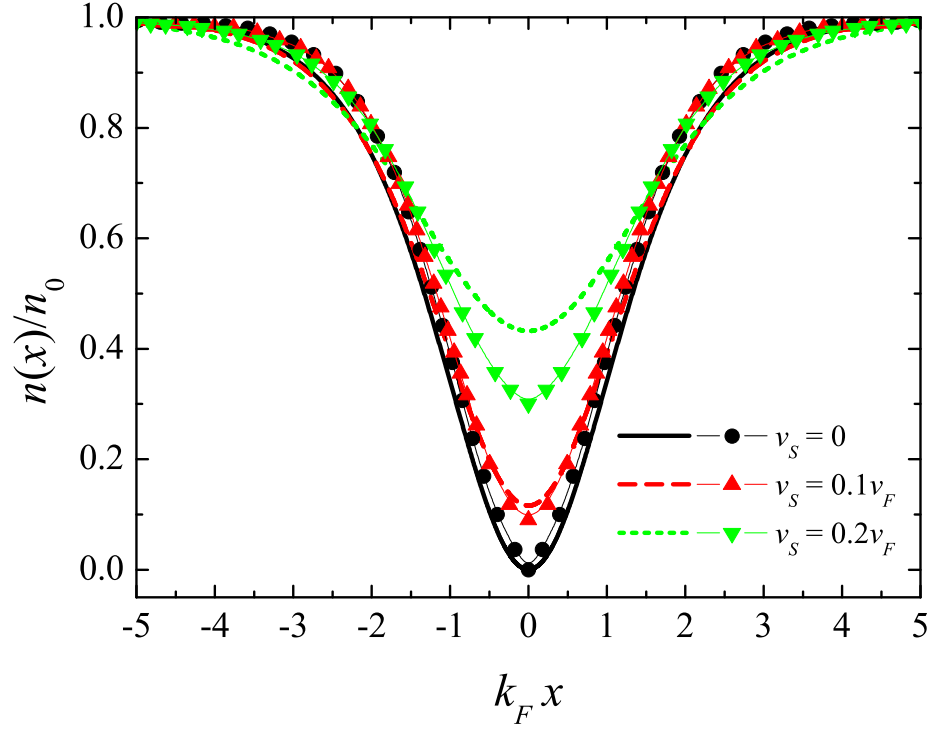}%
\caption{The fermion density profile $n_{x}/n_{\infty}$ of the dark soliton as
a function of distance to the soliton center (x=0) is shown for several
soliton velocities at $1/(k_{F}a_{s})=1$, and compared with BdG results
(dots).}%
\end{center}
\end{figure}

\newpage%

\begin{figure}
[h]
\begin{center}
\includegraphics[
height=6.5267in,
width=5.1327in
]%
{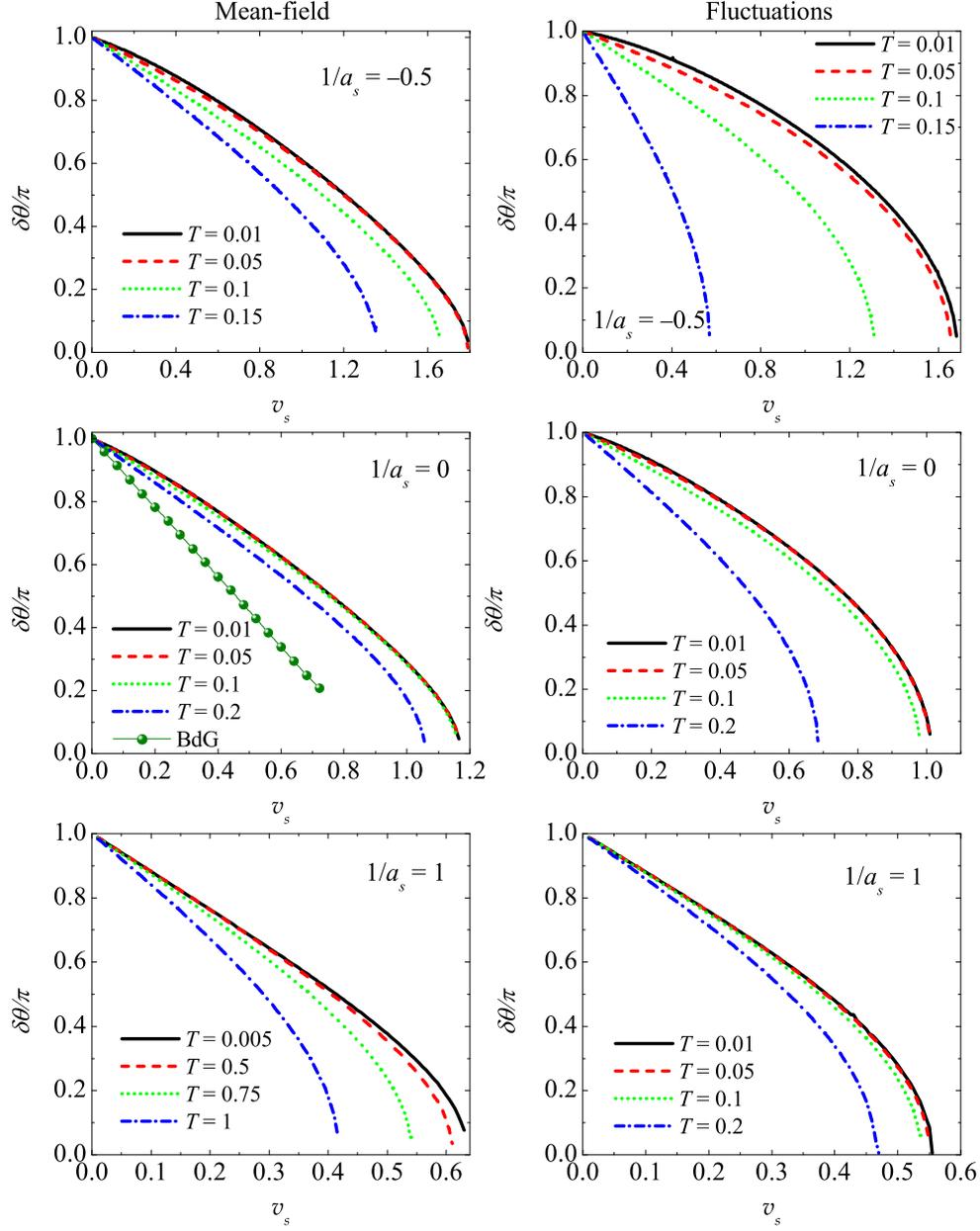}%
\caption{Phase difference $\delta\theta$ as a function on the soliton velocity
$v_{S}$. The notations are the same as in Figs. 1 and 2.}%
\end{center}
\end{figure}

\newpage%

\begin{figure}
[h]
\begin{center}
\includegraphics[
height=6.4714in,
width=5.2166in
]%
{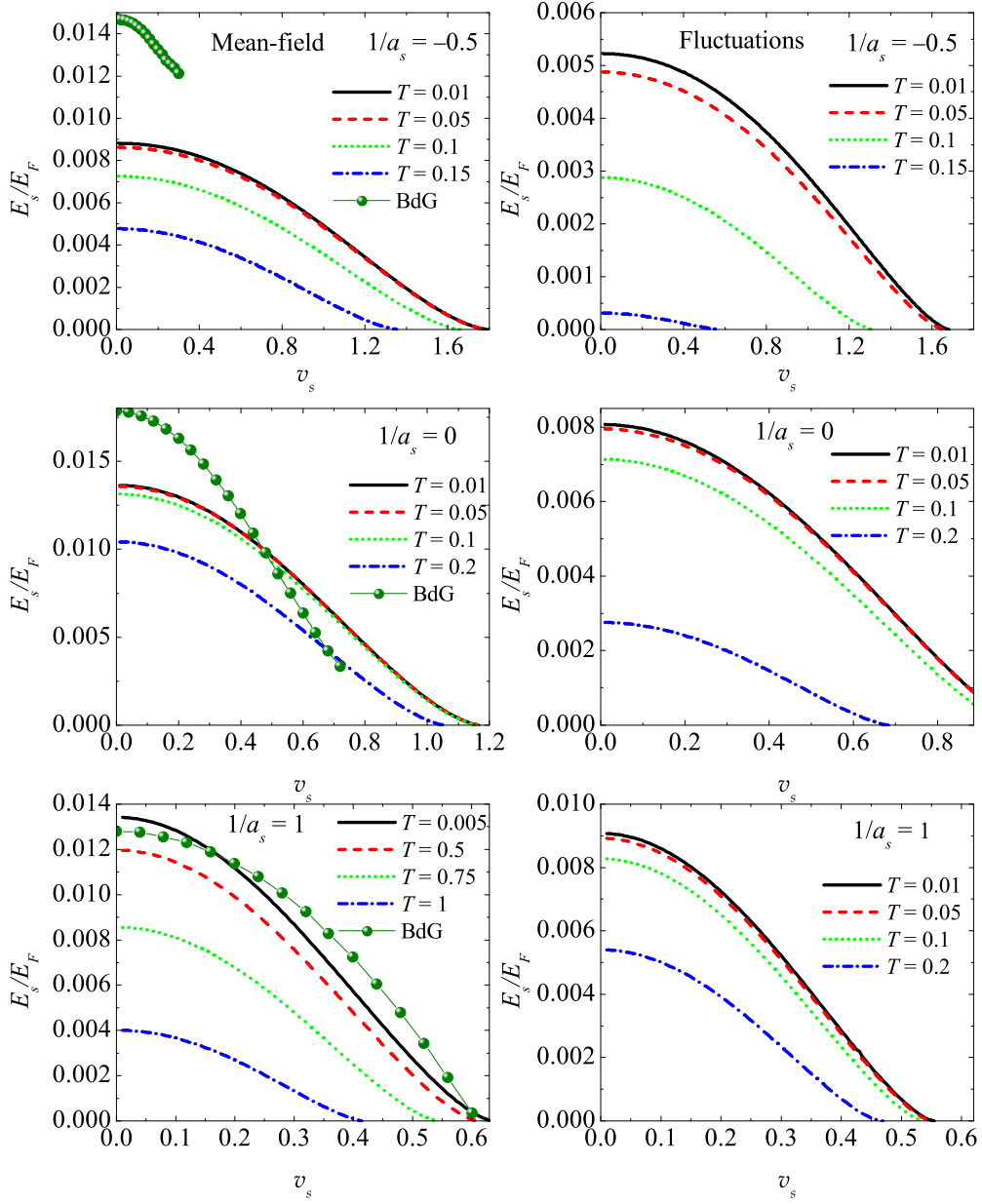}%
\caption{Soliton energy $\mathcal{E}_{S}\left(  v_{S}\right)  $ as a function
on the soliton velocity $v_{S}$. The full dots show the BdG data
\cite{Liao2011}.}%
\end{center}
\end{figure}

\newpage%

\begin{figure}
[h]
\begin{center}
\includegraphics[
height=6.429in,
width=5.214in
]%
{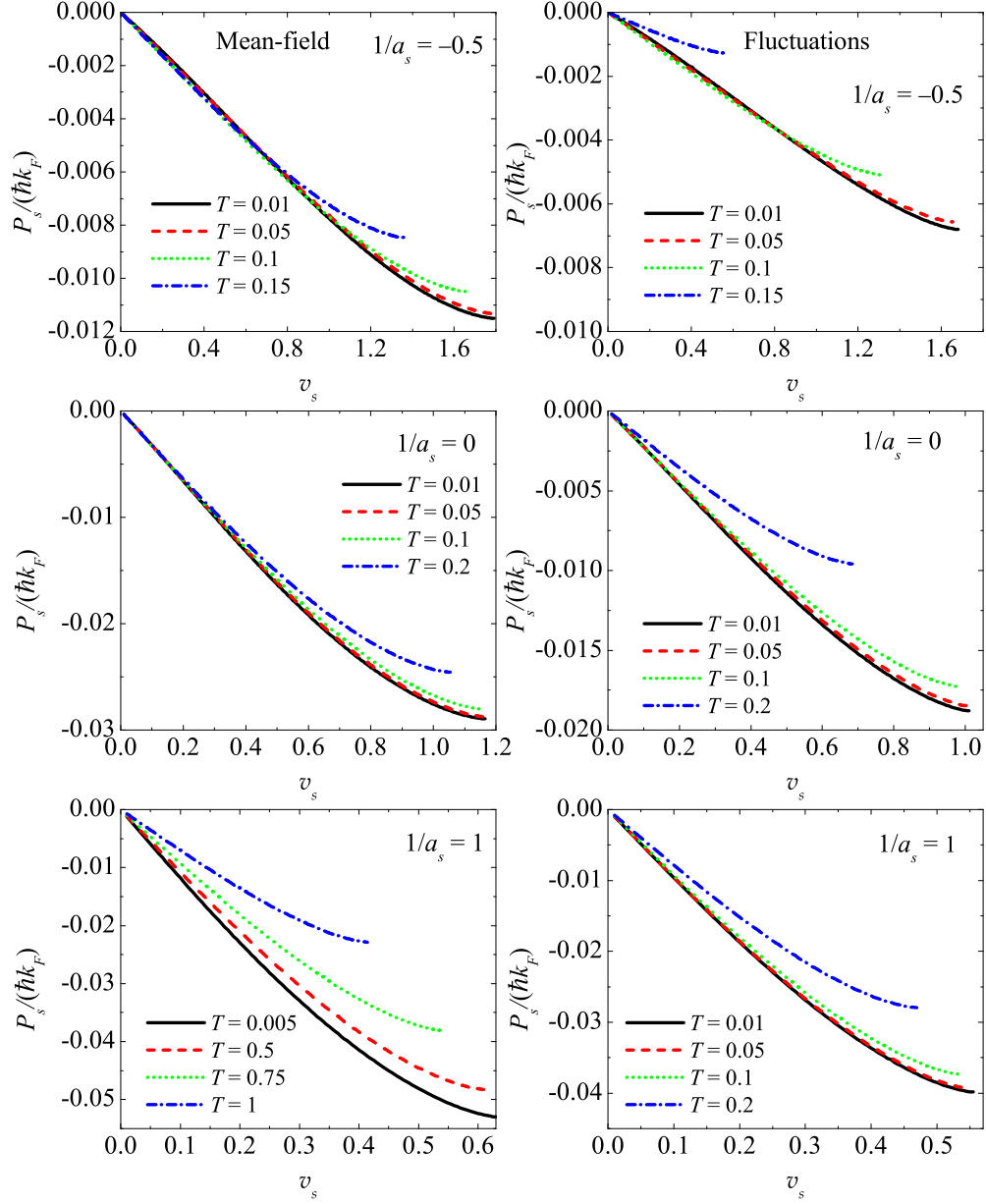}%
\caption{Soliton momentum $\mathcal{P}_{S}\left(  v_{S}\right)  $ as a
function on the soliton velocity $v_{S}$.}%
\end{center}
\end{figure}

\newpage%

\begin{figure}
[h]
\begin{center}
\includegraphics[
height=6.5414in,
width=5.2321in
]%
{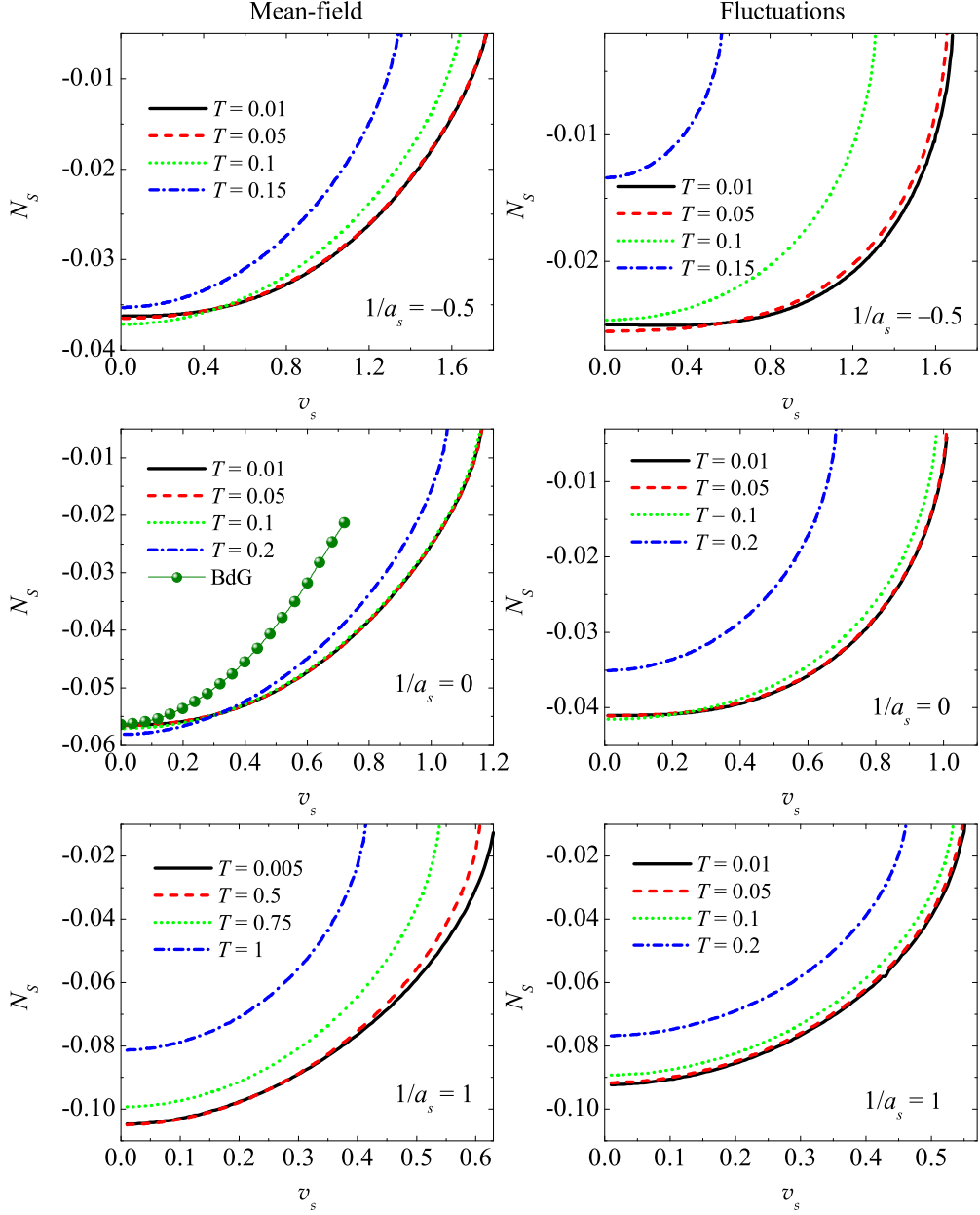}%
\caption{Number of fermions in the soliton $N_{S}\left(  v_{S}\right)  $ as a
function on the soliton velocity $v_{S}$. The full dots show the BdG results
of Ref. \cite{Liao2011}.}%
\end{center}
\end{figure}

\newpage%

\begin{figure}
[h]
\begin{center}
\includegraphics[
height=6.4013in,
width=5.188in
]%
{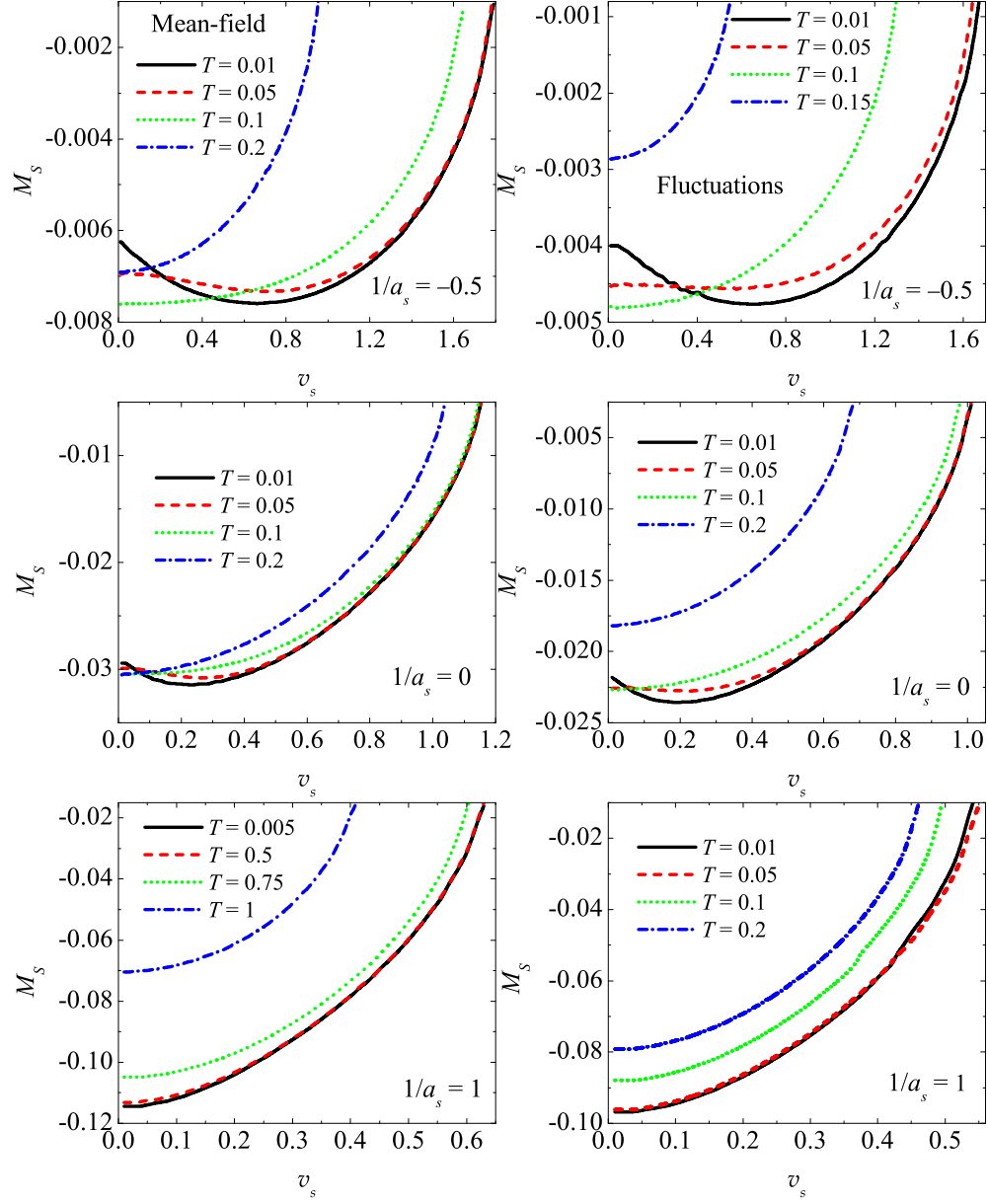}%
\caption{Effective mass of the soliton $M_{S}\left(  v_{S}\right)  $ as a
function on the soliton velocity $v_{S}$.}%
\end{center}
\end{figure}

\end{document}